\documentclass[aps,preprint]{revtex4}

\usepackage{makeidx}
\usepackage{amsmath}
\usepackage{graphicx}
\usepackage{epstopdf}
\usepackage{amssymb,epsf}

\begin{document}

\title{Quartic Quasi-Topological-Born-Infeld Gravity }

\author{Mohammad Ghanaatian}

\affiliation{Department of Physics, Payame Noor University, Iran}

\begin{abstract}
In this paper, quartic quasi-topological black holes in the
presence of a nonlinear electromagnetic Born-Infeld field is
presented. By using the metric parameters, the charged black hole
solutions of quasi-topological Born-Infeld gravity is considered.
The thermodynamics of these black holes are investigated and I
show that the thermodynamics and conserved quantities verify the
first law of thermodynamics. I also introduce the thermodynamics
of asymptotically AdS rotating black branes with flat horizon of
these class of solutions and I calculate the finite action by use
of the counterterm method inspired by AdS/CFT correspondence.
\end{abstract}

\maketitle


\section{Introduction}

The paradigm of extra dimensions does much more that excite our
imagination. It solves the so called "hierarchy of scales"
problem. Several extra-dimensional models have been introduced in
the past few years. It is well known that the natural
generalization of the Einstein-Hilbert action to higher
dimensional spacetime, and higher order gravity with second order
equation of motion, is the Lovelock action \cite{Lovelock}.
Because of the topological origin of the Lovelock terms, the
second term of the Lovelock action (the Gauss-Bonnet term) does
not have any dynamical effect in four dimensions. Similarly, the
cubic term just contributes to the equations of motion in seven
dimentions or greater. A modification of higher order Lovelock
gravity which contains cubic and quartic terms of Riemann tensor
and contributes to the equation of motions in five dimensions is
quasi-topological gravity \cite{Oliva,Myers1,Dehghani1}. In
second, third- \cite{Oliva,Myers1}, and fourth-order
\cite{Dehghani1} quasi-topological gravity, we deal with the field
equations of higher-derivative gravity in five and higher
dimensions. The equations of motion of these theories are
second-order differential equations for the spherically symmetric
metric. In order to the fact that the exact black hole solutions
have been constructed and also the equation of motion are only
second-order in derivatives in spherically symmetric setting,
contrary to the fact the quartic quasi-topological gravity is a
higher carvature theory of gravity, the holographic computations
in this model are still under control, at least for the spherical
symmetry. Furthermore, Since the quasi-topological theory contains
derivatives of metrics of order not higher than two (for the
spherical symmetry), the quantization of linearized
quasi-topological theory is free of ghosts. Thus, it is natural to
study the effects of these higher curvature terms on the
properties and thermodynamics of black holes and black branes
\cite{Brenna1,Dehghani2,Brenna2,Bazrafshan,Xiao1,Xiao2,Ghanaatian1,Ghanaatian2}.

\ \ It is well known that the rsh of gravitational equations (like
Einstein equation) is energy-momentum tensor, which relate to
matter and various fields. Among these fields the electromagnetic
field is so important. Without exaggeration the linear
electromagnetic field is one of the most successful theories of
electromagnetic fields. In order to solve the problem of
self-energy of electron, the theory of the non-linear
electromagnetic field was introduced by Born and Infeld (BI) in
1934 \cite{BI}. In the limits of weak fields, the Born-Infeld
lagrangian reduces to Maxwell lagrangian plus some small
corrections. If one is to consider the Maxwell fields coupled to a
gravitational action, which also includes string generated
corrections at higher orders, then, it is natural to consider
string generated corrections to the electromagnetic field action
as well. It is known that there are Born-Infeld terms which appear
as higher order corrections to the Maxwell action. In fact,
considering the analogy between the quasi-topological and the
Born-Infeld terms, it is worth to include both these corrections
simultaneously. In this letter, I will consider the quartic
quasi-topological gravity in the presence of nonlinear
electrmagnetic field and consider the black hole and black brane
solutions of this theory.

\ \ The outline of this paper is as follows: In Sec. \ref{Action},
a brief review of the quartic quasi-topological gravity in the
presence of a nonlinear Born-Infield electromagnetic field is
presented. In section \ref{Solutions}, I consider the charged
black holes of quasi-topological gravity in the presence of
Born-Infeld electromagnetic field. Section \ref{Black Hole} is
devoted to the investigation of the thermodynamic properties of
these solutions and the first law of thermodynamics. In Sec.
\ref{Black Brane}, I endow the solutions with asymptotically AdS
charged rotating black branes and study the thermodynamic
properties of them with flat horizon. Then, in Sec.
\ref{Conserved}, the finite action and conserved quantities of the
solutions are calculated. Finally, I finish my paper with some
concluding remarks in section \ref{Conclusion}.

\section{\ Quasi-Topological-Born-Infeld Action \label{Action}}

The action of quartic quasi-topological gravity in $(n+1)$
dimensions in the presence of a nonlinear Born-Infeld
electromagnetic field can be written as
\begin{equation}
I_{G}=\frac{1}{16\pi }\int d^{n+1}x\sqrt{-g}[-2\Lambda
+\mathcal{L}_{1}+\mu _{2}\mathcal{L}_{2}+\mu
_{3}\mathcal{X}_{3}+\mu _{4}\mathcal{X}_{4}+L(F)]. \label{Act1}
\end{equation}
where $\Lambda =-n(n-1)/2l^{2}$ is the cosmological constant, $\mathcal{L}_{1}={R}$ is just the Einstein-Hilbert Lagrangian, $%
\mathcal{L}_{2}=R_{abcd}{R}^{abcd}-4{R}_{ab}{R}^{ab}+{R}^{2}$ is
the second order Lovelock (Gauss-Bonnet) Lagrangian,
$\mathcal{X}_{3}$\ is the curvature-cubed Lagrangian \cite{Myers1}
\begin{eqnarray}
\mathcal{X}_{3} &=&R_{ab}^{cd}R_{cd}^{\,\,e\,\,\,f}R_{e\,\,f}^{\,\,a\,\,\,b}+%
\frac{1}{(2n-1)(n-3)}\left(
\frac{3(3n-5)}{8}R_{abcd}R^{abcd}R\right.
\notag \\
&&-3(n-1)R_{abcd}R^{abc}{}_{e}R^{de}+3(n+1)R_{abcd}R^{ac}R^{bd}  \notag \\
&&\left. +\,6(n-1)R_{a}{}^{b}R_{b}{}^{c}R_{c}{}^{a}-\frac{3(3n-1)}{2}%
R_{a}^{\,\,b}R_{b}^{\,\,a}R+\frac{3(n+1)}{8}R^{3}\right) .
\label{X3}
\end{eqnarray}
and $\mathcal{X}_{4}$ is the fourth order term of
quasi-topological gravity \cite{Dehghani1}
\begin{eqnarray}
\mathcal{X}_{4}\hspace{-0.2cm} &=&\hspace{-0.2cm}c_{1}R_{abcd}R^{cdef}R_{%
\phantom{hg}{ef}%
}^{hg}R_{hg}{}^{ab}+c_{2}R_{abcd}R^{abcd}R_{ef}R^{ef}+c_{3}RR_{ab}R^{ac}R_{c}{}^{b}+c_{4}(R_{abcd}R^{abcd})^{2}
\notag \\
&&\hspace{-0.1cm}%
+c_{5}R_{ab}R^{ac}R_{cd}R^{db}+c_{6}RR_{abcd}R^{ac}R^{db}+c_{7}R_{abcd}R^{ac}R^{be}R_{%
\phantom{d}{e}}^{d}+c_{8}R_{abcd}R^{acef}R_{\phantom{b}{e}}^{b}R_{%
\phantom{d}{f}}^{d}  \notag \\
&&\hspace{-0.1cm}%
+c_{9}R_{abcd}R^{ac}R_{ef}R^{bedf}+c_{10}R^{4}+c_{11}R^{2}R_{abcd}R^{abcd}+c_{12}R^{2}R_{ab}R^{ab}
\notag \\
&&\hspace{-0.1cm}%
+c_{13}R_{abcd}R^{abef}R_{ef}{}_{g}^{c}R^{dg}+c_{14}R_{abcd}R^{aecf}R_{gehf}R^{gbhd}.
\label{X4}
\end{eqnarray}%
with
\begin{eqnarray*}
c_{1} &=&-\left( n-1\right) \left( {n}^{7}-3\,{n}^{6}-29\,{n}^{5}+170\,{n}%
^{4}-349\,{n}^{3}+348\,{n}^{2}-180\,n+36\right) , \\
c_{2} &=&-4\,\left( n-3\right) \left( 2\,{n}^{6}-20\,{n}^{5}+65\,{n}^{4}-81\,%
{n}^{3}+13\,{n}^{2}+45\,n-18\right) , \\
c_{3} &=&-64\,\left( n-1\right) \left( 3\,{n}^{2}-8\,n+3\right) \left( {n}%
^{2}-3\,n+3\right) , \\
c_{4} &=&-{(n}^{8}-6\,{n}^{7}+12\,{n}^{6}-22\,{n}^{5}+114\,{n}^{4}-345\,{n}%
^{3}+468\,{n}^{2}-270\,n+54), \\
c_{5} &=&16\,\left( n-1\right) \left( 10\,{n}^{4}-51\,{n}^{3}+93\,{n}%
^{2}-72\,n+18\right) , \\
c_{6} &=&--32\,\left( n-1\right) ^{2}\left( n-3\right) ^{2}\left( 3\,{n}%
^{2}-8\,n+3\right) , \\
c_{7} &=&64\,\left( n-2\right) \left( n-1\right) ^{2}\left( 4\,{n}^{3}-18\,{n%
}^{2}+27\,n-9\right) , \\
c_{8} &=&-96\,\left( n-1\right) \left( n-2\right) \left( 2\,{n}^{4}-7\,{n}%
^{3}+4\,{n}^{2}+6\,n-3\right) , \\
c_{9} &=&16\left( n-1\right) ^{3}\left( 2\,{n}^{4}-26\,{n}^{3}+93\,{n}%
^{2}-117\,n+36\right) , \\
c_{10} &=&{n}^{5}-31\,{n}^{4}+168\,{n}^{3}-360\,{n}^{2}+330\,n-90, \\
c_{11} &=&2\,(6\,{n}^{6}-67\,{n}^{5}+311\,{n}^{4}-742\,{n}^{3}+936\,{n}%
^{2}-576\,n+126), \\
c_{12} &=&8\,{(}7\,{n}^{5}-47\,{n}^{4}+121\,{n}^{3}-141\,{n}^{2}+63\,n-9), \\
c_{13} &=&16\,n\left( n-1\right) \left( n-2\right) \left(
n-3\right) \left(
3\,{n}^{2}-8\,n+3\right) , \\
c_{14} &=&8\,\left( n-1\right) \left( {n}^{7}-4\,{n}^{6}-15\,{n}^{5}+122\,{n}%
^{4}-287\,{n}^{3}+297\,{n}^{2}-126\,n+18\right).
\end{eqnarray*}%
In the action (\ref{Act1}), $L(F)$  is the Born-Infeld Lagrangian
given as \cite{Hassaine1,Hassaine2,Maeda,Habib,Hendi1,Hendi2}

\begin{equation}
 L(F)=4\beta ^{2}\left( 1-\sqrt{1+\frac{F^{2}}{2\beta ^{2}}%
}\right) \label{LBI}.
\end{equation}
where $F=F_{\mu \nu }F^{\mu \nu }$, $F_{\mu \nu }=\partial _{\mu
}A_{\nu }-\partial _{\nu }A_{\mu }$ is the electromagnetic field
tensor and $A_{\mu } $ is the vector potential. One may note that
in the limit $\beta \longrightarrow\infty$ reduces to the standard
Maxwell form $L(F)=-F^{2}$.

\section{Charged Quasi-Topological-Born-Infeld Black Hole Solutions \label{Solutions}}

Now, I introduce the charged black hole solutions of
quasi-topological gravity in the presence of nonlinear Born-Infeld
electromagnetic field with Lagrangian (\ref{LBI}). The metric has
the following form:
\begin{equation}
ds^{2}=-f(\rho )dt^{2}+\frac{d\rho ^{2}}{f(\rho )}+\rho
^{2}d\Omega ^{2}.  \label{met0}
\end{equation}%
where
\[
d\Omega ^{2}=\left\{
\begin{array}{cc}
d\theta
_{1}^{2}+\sum\limits_{i=2}^{n-1}\prod\limits_{j=1}^{i-1}\sin
^{2}\theta _{j}d\theta _{i}^{2} & k=1 \\
d\theta _{1}^{2}+\sinh ^{2}\theta _{1}d\theta _{2}^{2}+\sinh
^{2}\theta _{1}\sum\limits_{i=3}^{n-1}\prod\limits_{j=2}^{i-1}\sin
^{2}\theta
_{j}d\theta _{i}^{2} & k=-1 \\
\sum\limits_{i=1}^{n-1}d\phi _{i}^{2} & k=0
\end{array}
\right\}\] represents the line element of an $(n-1)$-dimensional
hypersurface with constant curvature $(n-1)(n-2)k$ and volume
$V_{n-1}$. Using the metric (\ref{met0})  and
\begin{equation}
A_{\mu }=h(\rho )\delta _{\mu }^{0}.
\end{equation}%
for the vector potential, one can calculate the one dimensional
action after integration by parts. One obtains the action per unit
volume as
\begin{equation}
I_{G}=\frac{{(n-1)}}{16\pi l^{2}}\int dtd\rho \lbrack {{{\left[
\rho ^{n}(1+\psi +\hat{\mu}_{2}\psi ^{2}+\hat{\mu}_{3}\psi
^{3}+\hat{\mu}_{4}\psi
^{4})\right] ^{\prime }+\frac{4l^{2}\beta^{2}\rho ^{(n-1)}(1-\sqrt{1-\frac{h^{\prime 2}}{\beta^{2}})}}{%
(n-1)}}}}].  \label{Act3}
\end{equation}%
where $\psi =l^{2}\rho^{-2}(k-f)$ and the dimensionless parameters $%
\hat{\mu}_{2}$, $\hat{\mu}_{3}$ and $\hat{\mu}_{4}$ are defined
as:

\begin{equation*}
\hat{\mu}_{2}\equiv \frac{(n-2)(n-3)}{l^{2}}\mu _{2},\text{ \ \ \ }\hat{\mu}%
_{3}\equiv \frac{(n-2)(n-5)(3n^{2}-9n+4)}{8(2n-1)l^{4}}\mu _{3},
\end{equation*}%
\begin{equation*}
\hat{\mu}_{4}\equiv {\frac{n\left( n-1\right) \left( n-2\right)
^{2}\left(
n-3\right) \left( n-7\right) ({{n}^{5}-15\,{n}^{4}+72\,{n}^{3}-156\,{n}%
^{2}+150\,n-42)}}{{l}^{6}}}\mu _{4}.
\end{equation*}
 Variation with respect to $h(\rho )$ gives

\begin{equation}
(n-1) h^{\prime} ({\beta}^{2}- h^{\prime 2}) +\rho h^{''}
{\beta}^{2} =0, \label{eom2}
\end{equation}%
and therefore
 one can show that the vector potential can be written as
\begin{equation}
h(\rho)=-\sqrt{\frac{(n-1)}{2n-4}}\frac{q}{\rho^{n-2}} \Gamma(\eta
), \label{Amu0}
\end{equation}
where $q$ is an integration constant which is related to the
charge parameter and
\[
\eta =\frac{{(n-1)(n-2)q^{2}}}{2\beta ^{2}\rho^{2n-2}}.
\]
In Eq. (\ref{Amu0}) and throughout the paper, the following
abbreviation for the hypergeometric function is used,
\begin{equation}
{_{2}F_{1}}\left( {%
\left[ \frac{1}{2},\frac{n-2}{2n-2}\right] ,\left[
\frac{3n-4}{2n-2}\right] ,-z }\right) =\Gamma(z).  \label{hyp}
\end{equation}
The hypergeometric function $\Gamma(\eta ){\rightarrow 1}$ as
$\eta \rightarrow 0$ ($\beta \rightarrow \infty $) and therefore
$h(\rho)$ of Eq. (\ref{Amu0}) reduces to the gauge potential of
Maxwell field.

Varying the action (\ref{Act3}) with respect to $\psi (\rho)$
yields
\begin{equation}
\left( 1+2\hat{\mu}_{2}\psi +3\hat{\mu}_{3}\psi
^{2}+4\hat{\mu}_{4}\psi ^{3}\right) \frac{dN(\rho )}{d\rho }=0,
\label{eom1}
\end{equation}%
which shows that $N(\rho )$ should be a constant. Variation with
respect to $N(\rho )$ and substituting $N(\rho )=1$ gives
\begin{equation}
\hat{\mu}_{4}\psi ^{4}+\hat{\mu}_{3}\psi ^{3}+\hat{\mu}_{2}\psi
^{2}+\psi +\kappa =0,  \label{Eq4}
\end{equation}%
where
\begin{equation}
\kappa
=\hat{\mu}_{0}-\frac{m}{\rho^{n}}+\frac{4l^{2}\beta^{2}}{n(n-1)}[1-\sqrt{1+\eta}-\frac{\eta}{n-2}F(\eta)]
\end{equation}%
and $m$ is an integration constant which is related to the mass of
the spacetime. In order to obtain the black hole solutions, I
choose two solutions of $f(\rho)$ as
\begin{equation}
f_{1}(\rho)=k+\frac{\rho^{2}}{l^{2}}\left( \frac{\hat{\mu}_{3}}{4\hat{\mu}_{4}}+\frac{1%
}{2}R-\frac{1}{2}E\right) .  \label{Fr4}
\end{equation}
\begin{equation}
f_{2}(\rho)=k+\frac{\rho^{2}}{l^{2}}\left( \frac{\hat{\mu}_{3}}{4\hat{\mu}_{4}}-\frac{1%
}{2}R+\frac{1}{2}K\right) .  \label{Fr4}
\end{equation}

where
\begin{eqnarray}
R &=&\left( \frac{{\hat{\mu}_{3}}^{2}}{4{\hat{\mu}_{4}}^{2}}-\frac{\hat{\mu}%
_{2}}{\hat{\mu}_{4}}+y_{1}\right) ^{1/2},
\label{RR} \\
E &=&\left( \frac{3{\hat{\mu}_{3}}^{2}}{4{\hat{\mu}_{4}}^{2}}-\frac{2\hat{\mu%
}_{2}}{\hat{\mu}_{4}}-R^{2}-\frac{1}{4R}\left[ \frac{4\hat{\mu}_{2}\hat{\mu}%
_{3}}{{\hat{\mu}_{4}}^{2}}-\frac{8}{\hat{\mu}_{4}}-\frac{{\hat{\mu}_{3}}^{3}%
}{{\hat{\mu}_{4}}^{3}}\right] \right) ^{1/2},  \label{EE} \\
K &=&\left( \frac{3{\hat{\mu}_{3}}^{2}}{4{\hat{\mu}_{4}}^{2}}-\frac{2\hat{\mu%
}_{2}}{\hat{\mu}_{4}}-R^{2}+\frac{1}{4R}\left[ \frac{4\hat{\mu}_{2}\hat{\mu}%
_{3}}{{\hat{\mu}_{4}}^{2}}-\frac{8}{\hat{\mu}_{4}}-\frac{{\hat{\mu}_{3}}^{3}%
}{{\hat{\mu}_{4}}^{3}}\right] \right) ^{1/2}  \label{KK}\\
\Delta &=&\frac{H^{3}}{27}+\frac{D^{2}}{4},\text{ \ \ \ \ \ \ }H={\frac{3%
\hat{\mu}_{3}-{\hat{\mu}_{2}}^{2}}{3{\hat{\mu}_{4}}^{2}}}-\,{\frac{4\kappa }{%
\hat{\mu}_{4}},} \\
D &=&{\frac{2}{27}}\,{\frac{{\hat{\mu}_{2}}^{3}}{{\hat{\mu}_{4}}^{3}}}-\frac{%
1}{3}\,\left( {\frac{\hat{\mu}_{3}}{{\hat{\mu}_{4}}^{2}}}+8\,{\frac{\kappa }{%
\hat{\mu}_{4}}}\right) \frac{\hat{\mu}_{2}}{\hat{\mu}_{4}}+{\frac{{\hat{\mu}%
_{3}}^{2}\kappa
}{{\hat{\mu}_{4}}^{3}}}+\frac{1}{{\hat{\mu}_{4}}^{2}}.
\end{eqnarray}
and $y_{1}$ is the real root of following equation:
\begin{equation}
{y}^{3}-{\frac {\mu_{{2}}{y}^{2}}{\mu_{{4}}}}+ \left( {\frac
{\mu_{{3} }}{{\mu_{{4}}}^{2}}}-4\,{\frac {\kappa}{\mu_{{4}}}}
\right) y-{\frac {
{\mu_{{3}}}^{2}\kappa}{{\mu_{{4}}}^{3}}}+\,{\frac
{4\mu_{{2}}\kappa}{{ \mu_{{4}}}^{2}}}-\frac{1}{{\mu_{{4}}}^{2}}=0
\end{equation}

The metric function $f(\rho )$ for the uncharged solution $(q=0)$
is real in the whole range $0\leq \rho <\infty $. But for charged
solutions, one should restrict the spacetime to the region $\rho
\geq r_{0}$,  where $r_{0}$ is the largest real root of
 $\Delta_{0}=\Delta(\kappa=\kappa_{0})$,
 $R_{0}=R(\kappa=\kappa_{0})$,
 $E_{0}=E(\kappa=\kappa_{0})$ and
 $K_{0}=K(\kappa=\kappa_{0})$, and $\kappa_{0}$ is

\begin{equation}
\kappa_{0}
=\hat{\mu}_{0}-\frac{m}{r_{0}^{n}}+\frac{4l^{2}\beta^{2}}{n(n-1)}[1-\sqrt{1+\eta_{0}}-\frac{\eta_{0}}{n-2}F(\eta_{0})]
\end{equation}%

where
\[
\eta_{0} =\frac{{(n-1)(n-2)q^{2}}}{2\beta ^{2}{r_{0}}^{2n-2}}.
\]

Performing the transformation

\begin{equation}
r=\sqrt{{\rho }^{2}-{r_{0}}^{2}}\Rightarrow d{\rho }^{2}=\frac{r^{2}}{%
r^{2}+r_{0}^{2}}dr^{2}
\end{equation}%
the metric becomes%
\begin{equation}
ds^{2}=-f(r)dt^{2}+\frac{{r^{2}dr^{2}}}{(r^{2}+r_{0}^{2})f(r)}%
+(r^{2}+r_{0}^{2})\;\sum_{i=1}^{n-1}d\phi _{i}{}^{2}.
\end{equation}%
where now the functions $\eta$, $h(r)$ and $\kappa $ are
\[
\eta =\frac{{(n-1)(n-2)q^{2}}}{2\beta
^{2}{(r^{2}+r_{0}^{2})}^{{2n-2}/2}}.
\]

\begin{equation}
h(r)=-\sqrt{\frac{(n-1)}{2n-4}}\frac{q}{{(r^{2}+r_{0}^{2})}^{{(n-2)}/2}}
\Gamma(\eta ), \label{Amu1}
\end{equation}
\begin{equation}
\kappa
=\hat{\mu}_{0}-\frac{m}{{(r^{2}+r_{0}^{2})}^{n/2}}+\frac{4l^{2}\beta^{2}}{n(n-1)}[1-\sqrt{1+\eta}-\frac{\eta}{n-2}F(\eta)]
\end{equation}%

\section{Thermodynamics of Quasi-Topological-Born-Infeld Black Holes \label{Black Hole}}

One can obtain the Hawking temperature of the black hole solutions
that are considered  in the previous section as:
\begin{eqnarray}
T_+&=&\frac{f^{\prime }(r_{+})}{4\pi
}\sqrt{1+\frac{r_{0}^{2}}{r_{+}^{2}}} \notag
\\
&&={\frac{(n-1)[n\hat{\mu}%
_{0}\Upsilon_{+}^{8}+\left( n-2\right) k{l}^{2}\Upsilon_{+}^{6}+\left( n-4\right) {k}^{2}%
\hat{\mu}_{2}l^{4}\Upsilon_{+}^{4}+\left( n-6\right) k\hat{\mu}_{3}{l}%
^{6}\Upsilon_{+}^{2}+\left( n-8\right)
{k}^{2}\hat{\mu}_{4}{l}^{8}]}{\left(
\,\Upsilon_{+}^{6}+2k\hat{\mu}_{2}{l}^{2}\Upsilon_{+}^{4}\,+3k^{2}\hat{\mu}_{3}{l}%
^{4}\Upsilon_{+}^{2}\,+4\hat{\mu}_{4}k^{3}{l}^{6}\right) 4\pi(n-1)
\,{l}^{2}\Upsilon_{+}}} \notag
 \\&&+{\frac{4\Upsilon_{+}^{8}\beta
^{2}\left( 1-\sqrt{1+\Upsilon _{+}}\right)}{\left(
\,\Upsilon_{+}^{6}+2k\hat{\mu}_{2}{l}^{2}\Upsilon_{+}^{4}\,+3k^{2}\hat{\mu}_{3}{l}%
^{4}\Upsilon_{+}^{2}\,+4\hat{\mu}_{4}k^{3}{l}^{6}\right) 4\pi(n-1)
\,{l}^{2}\Upsilon_{+}}} \label{Temp}
\end{eqnarray}

where $\Upsilon_{+}=\sqrt{r_{+}^{2}+r_{0}^{2}}$ and $r_+$ is the
largest real root of $f(r)$.

The entropy density for black hole in quartic quasi-topological
gravity becomes,
\begin{equation}
S=\frac{{r_{+}^{n-1}}}{4}\left( 1+\,{\frac {2k \left( n-1 \right)
\hat{\mu}_{2}{l}^{2}}{
 \left( n-3 \right) r_{+}^{2}}}+\,{\frac {3{k}^{2} \left( n-1 \right) \hat{\mu}_{3}{
l}^{4}}{ \left( n-5 \right) r_{+}^{4}}}+\,{\frac
{4\hat{\mu}_{3}{k}^{4}{l}^{6}
 \left( n-1 \right) }{r_{+}^{6} \left( n-7 \right) }} \right) \label{Ent1}
\end{equation}

Calculating the flux of the electric field at infinity, one can
find the charge of the black hole as
\begin{equation}
Q=\frac{V_{n-1}}{4\pi }\sqrt{\frac{(n-1)(n-2)}{2}}q  \label{Ch}
\end{equation}

The electric potential $\Phi $ at infinity with respect to the
horizon can be defined by \cite{Cvetic,Caldarelli},
\begin{equation}
\Phi =A_{\mu }\chi ^{\mu }\left\vert _{r\rightarrow \infty
}-A_{\mu }\chi ^{\mu }\right\vert _{r=r_{+}}  \label{Pot1}
\end{equation}
where $\chi =\partial /\partial t$ is the null generator of the
horizon. The electric potential $\Phi $ can be found as follow:
\begin{equation}
\Phi
=\sqrt{\frac{(n-1)}{2(n-2)}}\frac{q}{{(r_{+}^{2}+r_{0}^{2})}^{(n-2)/2}}\Gamma
(\eta _{+}). \label{Pot2}
\end{equation}

By using the behavior of the metric at large $r$, the ADM
(Arnowitt-Deser-Misner) mass of black hole can be arrived. One can
easily show that the mass of the black hole is
\begin{equation}
M=\frac{V_{n-1}}{16\pi }\left( n-1\right) m.  \label{Mass1}
\end{equation}

In order to investigate the first law of thermodynamics, I use the
expression for the entropy, the charge, and the mass that are
given in Eqs. (\ref{Ent1}), (\ref{Ch}) and (\ref{Mass1}), and keep
$f(r_{+})=0$ in the mind, I introduce as

\begin{eqnarray}
M(S,Q) &=&\frac{\left( n-1\right){{(r_{+}^{2}+r_{0}^{2})}^{n/2}} }{16\pi }\left\{ \frac{4\beta^{2}}{n(n-1)}%
 \left[ 1-\sqrt{1+\Im }+\frac{(n-1)\Im }{(n-2)}\Gamma
(\Im )\right] + \right.  \nonumber \\
&&\left.  \hat{\mu}%
_{0}+k\frac{l^{2}}{{(r_{+}^{2}+r_{0}^{2})}}+\hat{\mu}_{2}{k}^{2}\frac{l^{4}}{{(r_{+}^{2}+r_{0}^{2})}^{2}}+%
\hat{\mu}_{3}{k}^{3}\frac{l^{6}}{{(r_{+}^{2}+r_{0}^{2})}^{3}}+\hat{\mu}_{4}{k}^{4}\frac{l^{8}}{%
{(r_{+}^{2}+r_{0}^{2})}^{4}}%
\right\},  \label{Sma}
\end{eqnarray}
where
\[
\Im =\frac{16\pi ^{2}Q^{2}}{\beta
^{2}{(r_{+}^{2}+r_{0}^{2})}^{n-1}}.
\]
In Eq. (\ref{Sma}), $r_{+}$\ is the real root of Eq. (\ref{Ent1})
which is a function of $S$. I can regard the parameters $S$ and
$Q$ as a complete set of extensive parameters for the mass
$M(S,Q)$ and define the intensive parameters $T$ and $\Phi$
conjugate to them. These quantities are the temperature and the
electric potential
\begin{equation}
T=\left( \frac{\partial M}{\partial S}\right) _{Q},\ \ \ \ \Phi
=\left( \frac{\partial M}{\partial Q}\right) _{S}.  \label{Dsma1}
\end{equation}
It is easy to show that the intensive quantities calculated by Eq.
(\ref{Dsma1}) that are obtained by computing $\partial M/\partial
r_{+}$ and $\partial S/\partial r_{+}$ and using the chain rule,
coincide with Eqs. (\ref{Temp}) and (\ref{Pot2}), respectively.
So, the thermodynamic quantities calculated in Eqs. (%
\ref{Temp}) and (\ref{Pot2}) lead to the first law of
thermodynamics,
\begin{equation}
dM=TdS+\Phi dQ . \label{1stlaw}
\end{equation}

\section{Thermodynamics of Charged Rotating Quasi-Topological-Born-Infeld Black Branes}
\label{Black Brane}

Now, I apply the spacetime solution (\ref{met0}) for $k=0$\ with a
global rotation. One may perform the following rotation boost in
the $t-\phi _{i}$ planes to add angular momentum to the spacetime
\begin{equation}
t\mapsto \Xi t-a_{i}\phi _{i},\hspace{0.5cm}\phi _{i}\mapsto \Xi \phi _{i}-%
\frac{a_{i}}{l^{2}}t  \label{Tr}
\end{equation}
where $[x]$ is the integer part of $x$ for $i=1...[n/2]$. The $SO(n)$ rotation group in $%
n+1$ dimensions shows the maximum number of rotation parameters.
So, the number of independent rotation parameters is $[n/2]$.
Therefore, for the flat horizon of the AdS rotating solution with
$p\leq \lbrack n/2]$, the metric can be written as follows
\cite{Awad}:
\begin{eqnarray}
ds^{2} &=&-f(r)\left( \Xi dt-{{\sum_{i=1}^{p}}}a_{i}d\phi _{i}\right) ^{2}+%
\frac{(r^{2}+r_{0}^{2})}{l^{4}}{{\sum_{i=1}^{p}}}\left(
a_{i}dt-\Xi l^{2}d\phi
_{i}\right) ^{2}  \nonumber \\
&&\ \text{ }+\frac{r^{2}dr^{2}}{(r^{2}+r_{0}^{2})f(r)}-\frac{(r^{2}+r_{0}^{2})}{l^{2}}{\sum_{i<j}^{p}}%
(a_{i}d\phi _{j}-a_{j}d\phi
_{i})^{2}+(r^{2}+r_{0}^{2}){{\sum_{i=p+1}^{n-1}}}d\phi _{i}.
\label{met1}
\end{eqnarray}
where $\Xi =\sqrt{1+\sum_{i}^{k}a_{i}^{2}/l^{2}}$. The vector
potential for this solution can be rewritten as
\begin{equation}
A_{\mu
}=-\sqrt{\frac{(n-1)}{2n-4}}\frac{q}{(r^{2}+r_{0}^{2})^{(n-2)/2}}\Gamma
(\eta )\left(
\Xi \delta _{\mu }^{0}-\delta _{\mu }^{i}a_{i}\right) \text{(no sum on }i%
\text{)}. \label{Amu}
\end{equation}

Analytic continuation of the metric, I can obtain the temperature
and angular momentum as follows:
\begin{eqnarray}
T&=&\frac{f^{\prime }(r_{+})}{4\pi\Xi
}\sqrt{1+\frac{r_{0}^{2}}{r_{+}^{2}}}.
\end{eqnarray}

\begin{equation}
\Omega _{i}=\frac{a_{i}}{\Xi l^{2}}.  \label{Om}
\end{equation}
Calculating the flux of the electric field at infinity, one drives
the electric charge per unit volume $V_{n-1}$:
\begin{equation}
Q=\frac{1}{4\pi }\sqrt{\frac{(n-1)(n-2)}{2}}\,\,\Xi q.
\label{Charge}
\end{equation}
According to the fact that $\chi =\partial _{t}+{\sum_{i}^{k}}%
\Omega _{i}\partial _{\phi _{i}}$ and using Eq. (\ref{Pot1}), the
electric potential $\Phi $ is obtained as
\begin{equation}
\Phi =\sqrt{\frac{(n-1)}{2(n-2)}}\frac{q}{\Xi
(r_{+}^{2}+r_{0}^{2})^{(n-2)/2}}\Gamma (\eta _{+}).  \label{Pot}
\end{equation}

\section{Conserved Quantities of the Solutions}
\label{Conserved}

The AdS/CFT correspondence mostly apply for infinite boundaries,
but sometimes it is used for finite boundaries in order to drive
the conserved and thermodynamic quantities
\cite{Dehghani5,Dehghani6}. Considering the thermodynamics of AdS
black holes by using the AdS/CFT correspondence, the deep insights
in the characteristics and phase structures of strong 't Hooft
coupling CFTs can be obtained. In this section, I drive the action
and conserved quantities of the solutions. In general, the action
and conserved quantities of the spacetime are divergent when
evaluated on the solutions. By using the counterterms method and
AdS/CFT correspondence, one can consider this divergence for
asymptotically AdS solutions of Einstein gravity \cite{Maldacena}.
The finite action for asymptotically AdS solutions with flat
boundary, $\widehat{R}_{abcd}(\gamma )=0$ may be written as
follow:
\begin{equation} \label{Actfinite}
I=I_{G}+I_{b}+I_{ct}
\end{equation}
In this finite action, boundary term is
\begin{equation}
I_{b}=I_{b}^{(1)}+I_{b}^{(2)}+I_{b}^{(3)}
\end{equation}
where, the following boundary term makes the Einstein-Hilbert
action well-defined \cite{Gibbons},
\begin{equation}
I_{b}^{(1)}=\frac{1}{8\pi }\int_{\partial
\mathcal{M}}d^{n}x\sqrt{-\gamma }K \label{Ib1}.
\end{equation}%
and the proper surface term for the Gauss-Bonnet term is
\cite{Myers2,Davis,Dehghani3,Dehghani4},
\begin{equation}
I_{b}^{(2)}=\frac{1}{8\pi }\int_{\partial \mathcal{M}}d^{n}x\sqrt{-\gamma }%
\left\{ \frac{2\hat{\mu _{2}}l^{2}}{(n-2)(n-3)}J\right\} .
\label{Ib2}
\end{equation}%
where $J$ is the trace of
\begin{equation}
J_{ab}=\frac{1}{3}%
(2KK_{ac}K_{b}^{c}+K_{cd}K^{cd}K_{ab}-2K_{ac}K^{cd}K_{db}-K^{2}K_{ab}).
\label{Jab}
\end{equation}%
and the surface terms for the curvature-cubed term of
quasi-topological gravity have been driven in Ref.
\cite{Dehghani2} as
\begin{eqnarray}
&&I_{b}^{(3)}=\frac{1}{8\pi }\int_{\partial \mathcal{M}}d^{n}x\sqrt{-\gamma }%
\Big\{\frac{3\hat{\mu _{3}}l^{4}}{5n(n-2)(n-1)^{2}(n-5)}%
(nK^{5}-2K^{3}K_{ab}K^{ab}  \notag \\
&&\,\ \ \ \ \ \ \ \ \ \ \ \ \ \ \ \ \ \ \ \ \ \ \ \ \ \ \ \
+4(n-1)K_{ab}K^{ab}K_{cd}K_{e}^{d}K^{ec}-  \notag \\
&&\,\ \ \ \ \ \ \ \ \ \ \ \ \ \ \ \ \ \ \ \ \ \ \ \ \ \ \ \
(5n-6)KK_{ab}[nK^{ab}K^{cd}K_{cd}-(n-1)K^{ac}K^{bd}K_{cd}])\Big\}.
\label{Ib3}
\end{eqnarray}
and
\begin{eqnarray}
&&I_{b}^{(4)}={\frac{1}{8\pi }}\int_{\partial \mathcal{M}}d^{n}x\sqrt{%
-\gamma }{\frac{\hat{2\mu_{4}}{l}^{6}}{7n(n-1)\left( n-2\right)
\left( n-7\right) \left( {n}^{2}-3\,n+3\right) }}\Big\{\alpha
_{1}K^{3}K^{ab}K_{ac}K_{bd}K^{cd}  \notag \\
&&\,\ \ \ \ \ \ \ \ \ \ \ \ \ \ \ \ \ \ \ \ \ \ \ \ \ \ \ \
+\alpha _{2}K^{2}K^{ab}K_{ab}K^{cd}K_{c}^{e}K_{de}+\alpha
_{3}K^{2}K^{ab}K_{ac}K_{bd}K^{ce}K_{e}^{d}  \notag \\
&&\,\ \ \ \ \ \ \ \ \ \ \ \ \ \ \ \ \ \ \ \ \ \ \ \ \ \ \ \
+\alpha _{4}KK^{ab}K_{ab}K^{cd}K_{c}^{e}K_{d}^{f}K_{ef}+\alpha
_{5}KK^{ab}K_{a}^{c}K_{bc}K^{de}K_{d}^{f}K_{ef}  \notag \\
&&\,\ \ \ \ \ \ \ \ \ \ \ \ \ \ \ \ \ \ \ \ \ \ \ \ \ \ \ \
+\alpha _{6}KK^{ab}K_{ac}K_{bd}K^{ce}K^{df}K_{ef}+\alpha
_{7}K^{ab}K_{a}^{c}K_{bc}K^{de}K_{df}K_{eg}K^{fg}\Big\} .
\label{Ib4}
\end{eqnarray}%
presents a surface term which makes the action of quartic
quasi-topological gravity well-defined \cite{Bazrafshan}. And the
last term, $I_{ct}$, which is a functional of the boundary
curvature invariants, is counterterm and may be written as follow
\cite{Henningson,Nojiri,Hyun,Bala}:
\begin{equation}
I_{ct}=-\frac{1}{8\pi }\int_{\partial \mathcal{M}}d^{n}x\sqrt{-\gamma }\;%
\frac{(n-1)}{l_{eff}}.  \label{Ict}
\end{equation}%
where $l_{eff}$ is a scale length factor which depends on $l$ and
coupling constants of gravity and should reduce to $l$ in the
absence of higher curvature term. By using this counterterm
method, the action (\ref{Actfinite}) may be finite and can be used
it to calculate the conserved and thermodynamics quantities.

The conserved quantities associated with the Killing vectors
$\partial /\partial t$ and $\partial /\partial \phi ^{i}$ can be
obtained as
\begin{eqnarray}
M &=&\frac{1}{16\pi }m\left( n\Xi ^{2}-1\right) .  \label{Mass} \\
J_{i} &=&\frac{1}{16\pi }n\Xi ma_{i} .  \label{Angmom}
\end{eqnarray}
which are the mass and angular momentum of the solution.

By using Gibbs-Duhem relation
\begin{equation}
S=\frac{1}{T}(M-Q\Phi -{{\sum_{i=1}^{k}}}\Omega _{i}J_{i})-I,
\label{GibsDuh}
\end{equation}
one can introduce the entropy per unit volume $V_{n-1}$ as follow:
\begin{equation}
S=\frac{\Xi }{4}r_{+}^{n-1}  \label{Entropy}
\end{equation}
This shows that the entropy obeys the area law for our case where
the horizon is flat.

\section{Concluding Remarks \label{Conclusion}}

In this paper, I introduced the quartic quasi-topological gravity
in the presence of Born-Infeld field which is a nonlinear
electromagnetic field. I computed the charged black hole solutions
of this theory. These solutions presented a black holes with one
or two horizons or a naked singularity depending on the values of
charge and mass parameters. I investigated that the solutions
reduce to the solutions of Einstein-Born-Infeld when the the
Gauss-Bonnet and quasi-topological coefficients ($\mu_2$, $\mu_3$
and $\mu_4$) vanish, and reduce to the solutions of quartic
quasi-topological gravity in the presence of Maxwell field as
$\beta$ goes to infinity. In order to verify the first law of
thermodynamics of these black hole solutions, I calculated the
thermodynamic quantities $S$, $Q$ and $M$ where the mass was
driven as a function of the extensive parameters $S$ and $Q$.

Then, I considered the thermodynamics of asymptotically AdS black
branes with a flat horizon. I used the boundary term to find the
on-shell action of the quartic quasi-topological gravity. However,
one can use this method in the Hamiltonian formalism of
quasi-topological gravity. As in the case of Einstein solutions,
the action and the conserved quantities of the quasi-topological
solutions are not finite, I used the counterterm method to drive
the finite action and the conserved quantities. The counterterm
method inspired by the anti-de Sitter conformal field theory
(AdS/CFT) correspondence has been widely applied to the case of
Einstein gravity. The conserved quantities of black branes of
quasi-topological gravity with a flat horizon are independent the
coefficients $\mu_2$, $\mu_3$ and $\mu_4$ for fixed values of the
metric parameters, mass, charge, and rotation parameter. But, I
found that the thermodynamic quantities, the temperature, and the
entropy depend on the coupling constants $\mu_2$, $\mu_3$ and
$\mu_4$ through the value of the radius of the horizon. By using
the Gibbs-Duhem relation, I found the entropy and presented that
the entropy obeys the area law for black branes with flat horizon.

\acknowledgements This work has been supported by Payame Noor
University.

\end{document}